\documentclass[fleqn,usenatbib]{mnras}

\usepackage{newtxtext,newtxmath}

\usepackage[T1]{fontenc}
\usepackage{ae,aecompl}

\usepackage{graphicx}	
\usepackage{amsmath}	
\usepackage{amssymb}	
\usepackage[normalem]{ulem}

\newcommand{\rev}[1]{{#1}}

\newcommand{\aaao}[1]{}

\newcommand{\pppo}[1]{}

\newcommand{\jjjo}[1]{}

\title[Pebble dynamics and accretion onto rocky planets.]{Pebble dynamics and accretion onto rocky planets.\\ II. Radiative models}

\author[{A. Popovas, \AA}. Nordlund, J. P. Ramsey]{
Andrius Popovas$^{1,2}$\thanks{E-mail: popovas@nbi.ku.dk (AP)},
{\AA}ke Nordlund$^{1}$\thanks{E-mail: aake@nbi.ku.dk (\AA N)}
and Jon P.\ Ramsey$^{1,3 }$\thanks{E-mail: jpramsey@virginia.edu (JPR)}
\\
$^{1}$Center for Star and Planet Formation, the Niels Bohr Institute and the Natural History Museum of Denmark,\\
University of Copenhagen, \O ster Voldgade 5-7, DK-1350 Copenhagen, Denmark\\
$^{2}$Rosseland Centre for Solar Physics, Institute of Theoretical Astrophysics, 
University of Oslo, P.O. Box 1029 Blindern, N-0315 Oslo, Norway\\
$^{3}$Department of Astronomy, University of Virginia, Charlottesville, VA 22904, USA
}

\date{Accepted 2018 October 16. Received 2018 October 3; in original form 2018 July 20}

\pubyear{2018}

\begin{document}
\label{firstpage}
\pagerange{\pageref{firstpage}--\pageref{lastpage}}
\maketitle



\begin{abstract}
We investigate the effects of radiative energy transfer on a series of nested-grid, high-resolution hydrodynamic simulations of gas and particle dynamics in the vicinity of an Earth-mass planetary embryo.  We include heating due to the accretion of solids and the subsequent convective motions. Using a constant embryo surface temperature, we show that radiative energy transport results in a tendency to reduce the entropy in the primordial atmosphere, but this tendency is alleviated by an increase in the strength of convective energy transport, triggered by a correspondingly increased super-adiabatic temperature gradient. \rev{As a consequence, the amplitude of the convective motions increase by roughly an order of magnitude in the vicinity of the embryo.} In the cases investigated here, where the optical depth towards the disk surface is larger than unity, the reduction of the temperature in the outer parts of the Hill sphere relative to cases without radiative energy transport is only $\sim$100K, while the mass density increase is on the order of a factor of two in the inner parts of the Hill sphere. Our results demonstrate that, unless unrealistically low dust opacities are assumed, radiative cooling in the context of primordial rocky planet atmospheres can only become important after the disk surface density has dropped significantly below minimum-mass-solar-nebula values.
%
%
\end{abstract}


\section{Introduction}

Planetary embryos of sufficient mass embedded in optically thick protoplanetary accretion disks are necessarily accompanied by extended primordial atmospheres in near hydrostatic equilibrium, merging smoothly with the disk background at distances on the order of the Hill radius, $R_\mathrm{H} = a\sqrt[3]{\frac{M_\mathrm{p}}{3 M_*}}$, where $a$ is the semi-major axis of the embryo's orbit, and $M_\mathrm{p}$ and $M_*$ are the masses of the embryo and the central star, respectively \citep{perricameron1974, bodenheimerpollack1986, Pollack1996, dangelo2013, alibert_maximum_2017}.

The existence of dust and solid particles in the protoplanetary disk leads to accretion of solids onto such embryos, resulting in growth of the embryo, accretion heating at the embryo surface, and frictional heating of the surrounding primordial atmosphere \citep[e.g.][]{Brouwers2018}. As long as the embryo atmosphere and surrounding disk remain optically thick, convective motions are the most effective means of transporting heat through the atmosphere, with efficient convective transport leading to a near-adiabatic stratification of the atmosphere \citep{stevenson1982_giantplanets,wuchterl1993_envelopeconvection}. Near-adiabaticity implies that the inner parts of the atmosphere are relatively hot, which, when combined with the reduction of the embryo's gravity as the square of the distance, results in extended, relatively low mass atmospheres \citep[e.g.][]{venturinietal2015}.

If, on the other hand, the optical depth of the primordial atmosphere is not very large, radiative energy transport leads to cooling, resulting in possibly significant changes to the stratification of the atmosphere. Specifically, radiative cooling would lead to smaller pressures and density scale heights, thus increasing the mass of the atmosphere. Significant cooling would eventually lead to a situation where hydrostatic equilibrium is no longer sustainable, resulting in gravitational collapse and the formation of gas giants \citep[e.g.][]{mizunoetal1978, bodenheimerpollack1986, horiikoma2011}.  Opposing this effect is the replenishment of gas inside the Hill sphere with disk material \citep{Ormel2015b,Cimerman2017}.

In cases where catastrophic collapse of the atmosphere does not occur, the ultimate formation outcome depends crucially on the balance between loss of atmospheric gas caused by the reduction of the density and pressure of the surrounding disk over time, and the opposing tendency caused by the increased radiative cooling---occurring for essentially the same reason.  As shown by \citet{Ginzburg2016}, this balance may, in the end, dictate if the final outcome is a \rev{gas-rich planet}, or a rocky planet with a thin remnant atmosphere.

Properly accounting for the radiative transfer (RT) of energy and the corresponding effects on the near-hydrostatic equilibrium of primordial atmospheres is thus crucial for realistic and accurate modelling of early planet formation. Such efforts are additionally complicated by effects of scattering, which are known to be important at the wavelengths and temperatures of relevance \citep[e.g.][]{pinteetal2009_diskbench}.

In this Letter, we present first-of-a-kind results, using ray tracing RT with scattering to demonstrate its effects on primordial atmospheres, and the resulting effects on particle dynamics and pebble accretion onto rocky planets.
%
\section{Methods}
\label{sec:methods}
As in \citet[hereafter PNRO18]{popovas2018}, this study is carried out using the DISPATCH framework \citep{Nordlund2018}, employing a three-dimensional, Cartesian (shearing box) domain, with a set of static, nested patches. We continue to use an ideal gas equation of state (EOS) with adiabatic index $\gamma$ = 1.4 and molecular weight $\mu$ = 2, and embedment in a disk with a nominal surface mass density of 170 g cm$^{-2}$, corresponding to 1/10 of the minimum-mass-solar-nebula (MMSN). The grid set up is the same as in PNRO18 for a \rev{$M_{\rm p}=0.95 M_\oplus$} planet at 1 AU distance from the central star. We conduct a series of radiative-convective simulations, for which the basic RT and accretion heating parameters are summarized in Table \ref{tab:initial_conditions}. 

\rev{To investigate the effects of radiative cooling while maintaining a nearly adiabatic (i.e. convective) atmosphere, we deliberately choose a combination of disk surface density and opacity such that the disk and embryo atmosphere are slightly optically thick.} To this end we adopt a total opacity (with 80\% scattering and 20\% absorption) of 0.1 cm$^2$\,g$^{-1}$, and a disk surface density of 170 g\,cm$^{-2}$. We thus obtain a midplane optical depth in the unperturbed disk of $\frac{1}{2} \times 0.1 \times 1700/10 = 8.5$. The additional optical depth of the initial, adiabatic atmosphere is $\approx 6.0$, and hence the optical properties are such that we expect to begin to see the effects of radiative cooling.

\rev{The optical properties of dust and gas in hot, primordial atmospheres are quite uncertain, mainly due to uncertainties related to dust coagulation and thermal processing. Nevertheless, if we consider the opacities from \citet{Semenovetal2003} at densities and temperatures relevant to this work (e.g.\ Fig.\ \ref{fig:temp_rho_shell}), then the vast majority of our domain has opacities on the order of 0.1 -- 1 cm$^2$\,g$^{-1}$.}
%
\subsection{Initial and boundary conditions}
For initial conditions we take a fiducial, fully relaxed MMSN/10 run with adiabatic stratification (\texttt{m095t10} in PNRO18). The external and spherical boundary conditions for density $\rho$, entropy per unit mass and mass flux are also the same as in PNRO18. However, as we also consider  \textbf{}radiative energy transport in the current work, appropriate boundary conditions must be considered. The disk is optically thick in the radial and azimuthal directions and is initially in radiative equilibrium in those directions, i.e., no heat exchange occurs radially or azimuthally. However, protoplanetary disks do cool radiatively in the vertical direction. Therefore, we adopt the following external boundary conditions for the radiative transfer: \rev{$Q = I - S = 0$ at external horizontal (radial and azimuthal) boundaries; $I^{\rm incoming} = 0 \Rightarrow Q^{\rm incoming} = -S$ at external vertical boundaries, where $I$ is the radiation intensity in a specific direction, $S$ is the local source function, and $Q$ is the heating/cooling rate.}
\begin{table}
\centering
{%
\caption{\label{tab:initial_conditions} Simulation parameters. $\kappa$ is the total opacity (absorption plus scattering). $\epsilon$ is the fraction of absorption. $\dot{M}$ is the solid accretion rate adopted for the embryo heating term.}
\begin{tabular}{lccc}
Run     & $\kappa$ [cm$^2$\! g$^{-1}$]  & $\epsilon$ & $\dot{M}$ [M$_\oplus$\! yr$^{-1}$]  \\
\hline
\texttt{m095}               & --    & --   & --             \\
\texttt{m095-conv-2e-6}     & --    & --   & 2\ 10$^{-6}$ \\
\texttt{m095-conv-2e-6-rt}  & 0.1  & 0.2 & 2\ 10$^{-6}$ \\
\texttt{m095-conv-2e-5-rt}  & 0.1  & 0.2 & 2\ 10$^{-5}$ \\
\texttt{m095-conv-2e-6-rt-$\kappa$1}  & 1.0  & 0.2 & 2\ 10$^{-6}$
\end{tabular}}
\end{table}
%
\subsubsection{Heating from the embryo}
\rev{Accretion heating due to solids is included via a source term in the entropy equation (PNRO18), $Q_{\rm accr} \rho / P_{\rm gas}$, where $Q_{\rm accr} = \dot{M}GM_{\rm p} / 4\pi r^4$, $r$ is the distance from the embryo, $G$ is the gravitational constant and $\dot{M}$ is the adopted accretion rate (Tab.\ \ref{tab:initial_conditions}).}

As the envelope surrounding the embryo is optically thick \rev{for} most of the disk evolution, the embryo and its atmosphere cannot radiate heat away effectively when solids are accreted. The atmosphere is thus expected to remain nearly adiabatic through most of the build-up of the embryo mass, and the embryo must be correspondingly hot, especially towards the end of the build-up period. Detailed predictions of \rev{the temperature and its history requires} realistic equations of state for both the primordial atmosphere and the planet, in combination with modelling of the \rev{embryo} thermal evolution over disk evolution time scales. This is a formidable task by itself, where sequences of simulations such as the ones reported here may be used to provide the required estimates of instantaneous cooling rates.

In any case, since the heat capacity of a planet is large, the planet is expected to remain hot for a considerable time, even after the envelope becomes optically thin \citep{Ginzburg2016}. For the relatively brief periods of time covered by the current simulations, it is thus appropriate to consider the surface temperature of the embryo to be fixed, and we therefore adopt the surface temperature predicted by the adiabatic atmosphere model as a fixed lower boundary condition in all models.

\subsection{Radiative energy transport}
We use a hybrid-characteristics ray tracing scheme \citep[][Appendix \ref{app:ray_tracing}, available online only]{Nordlund2018} with 26 ray-directions (forward and reverse directions along 3 axes, 6 face diagonals, and 4 space diagonals) and a single frequency bin (constant opacity $\kappa$).  Since scattering is an important effect at the temperatures and wavelengths of relevance here, we split the opacity into an absorption part, $\epsilon \kappa$, and a scattering part, $(1-\epsilon) \kappa$.  At the levels of scattering typically assumed---from 50\% to 80\%---and with the relatively slow time evolution dictated by convective motions, scattering can be handled with what is essentially ``$\Lambda$ iteration'', i.e., iteratively feeding the mean intensity from previous time steps into the source function of the next time step \citep{Hubeny2003}. Here, the time slices \rev{stored} in DISPATCH may be used to \textit{predict} the mean intensity in the next time step, thus efficiently reducing the time lag otherwise resulting from the use of the previous mean intensity.

\begin{figure}
\centering
    \includegraphics[width=\columnwidth,clip,trim=0 35 0 10]{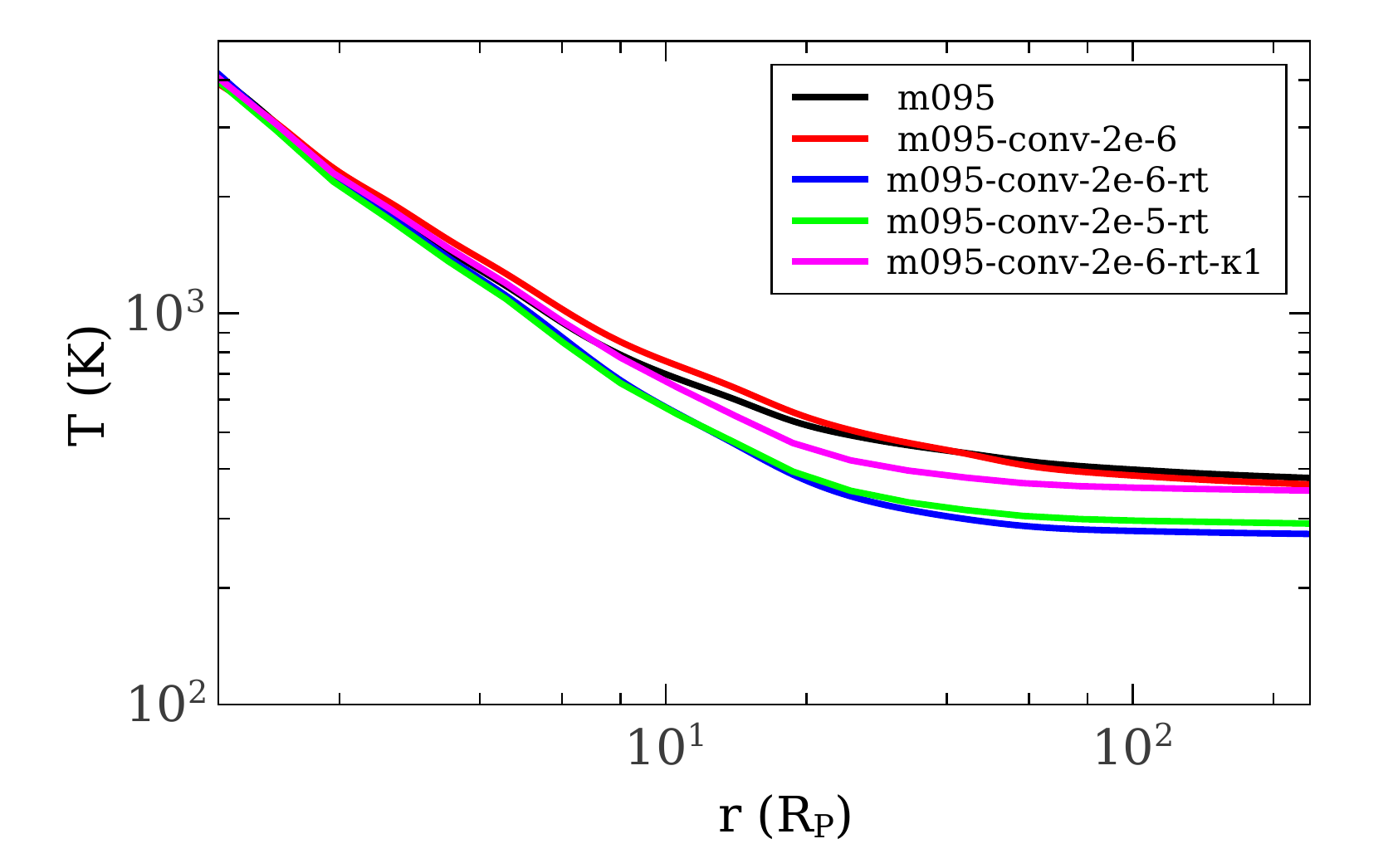}\\[-1.0ex]
    \includegraphics[width=\columnwidth,clip,trim=0 10 0 0]{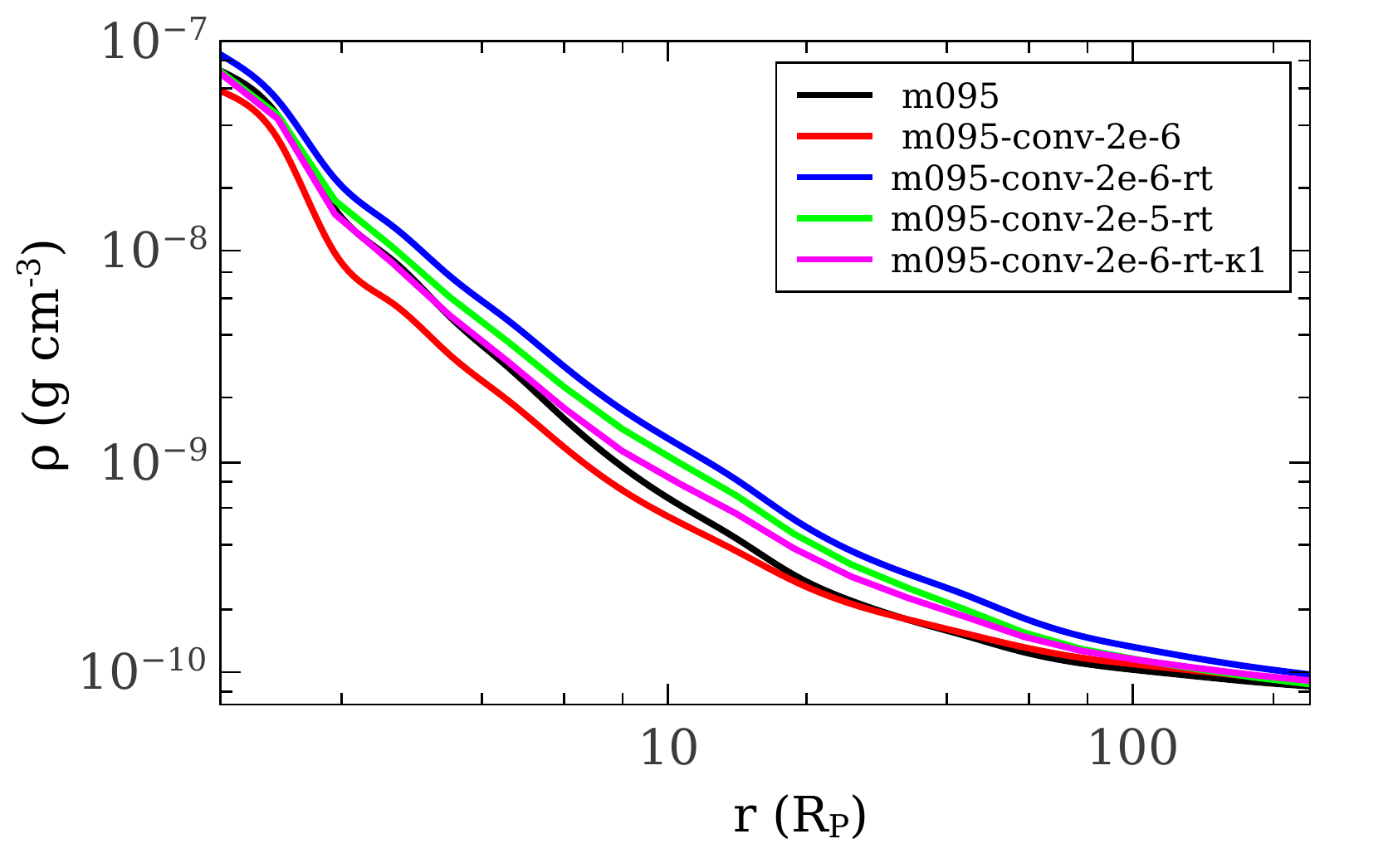}
    \caption{Temperature and density profiles, averaged over radial shells and time \rev{(see PNRO18)}, when neither RT, nor accretion heating is considered (\texttt{m095}, black curves), when accretion heating ($\dot{M} = 2\ 10^{-6}$ M$_{\oplus}$\! yr$^{-1}$) and driven convection but no RT is considered (\texttt{m095-conv-2e-6}, red), when accretion heating, convection and RT (with $\kappa$=0.1 cm$^2$\! g$^{-1}$) is considered (\texttt{m095-conv-2e-6-rt}, blue), when the same accretion heating and radiative cooling is considered, but $\kappa$=1.0 cm$^2$\! g$^{-1}$ (\texttt{m095-conv-2e-6-rt-$\kappa$1}, magenta), and when the accretion heating is increased to rates consistent with pebble-size particle accretion ($\dot{M} = 2 \ 10^{-5}$ M$_{\oplus}$\! yr$^{-1}$) and RT (with $\kappa$=0.1 cm$^2$\! g$^{-1}$) is considered (\texttt{m095-conv-2e-5-rt}, green).}
    \label{fig:temp_rho_shell}
\end{figure}

In order to avoid that the intense radiation from the embryo surface results in narrow, parallel beams of strong radiation in the 26 angular directions used in the ray-based radiation solver, we replace \rev{the solution from} the ray-based solver with a diffusion approximation in the region near the embryo. In the optically thick limit, the integral solution of the radiative transfer equation along any direction may be approximated by
\begin{equation}
Q = \frac{1}{2}(I^+ + I^- ) - S \approx \frac{d^2S}{d \tau^2} ,
\label{qdiffusion}
\end{equation}
where $I^+$ and $I^-$ are the specific radiation intensities in the forward and reverse directions, and $\tau$ is the optical depth along the ray direction.
A term proportional to the first derivative of the source function is omitted, since it cancels out when averaging forward and reverse solutions along a given ray direction.  Given the availability of the source function and opacity values on the Cartesian mesh, it is trivial to evaluate this expression in the three axis directions, and estimate the full space angle, integrated heat exchange rate per unit mass:
\begin{equation}
q \approx \kappa\frac{4\pi}{3} \left(\frac{d^2 S}{d\tau_x^2}+\frac{d^2 S}{d\tau_y^2}+\frac{d^2 S}{d\tau_z^2}\right) .
\label{qdiffuse}
\end{equation}
\rev{The heat exchange rate for the ray-based radiation solver, after averaging over the 26 forward and reverse ray directions, is meanwhile:
\begin{equation}
q \approx \kappa\frac{4\pi}{26} \sum_{j=1}^{13} (Q_j^+ + Q_j^-) .
\label{eq:rt_qheating}
\end{equation}
We apply the ray-based solver everywhere, but we weight the resultant heat exchange rate by
\begin{equation}
w=e^{-(r/r_0)^6},
\label{eq:rt_switch_weight}
\end{equation}
and the diffusion-based value of $q$ from Eq.\ \ref{qdiffuse} by $(1-w)$, where $r_0$ is 10 times the radius of the embryo.}

\subsection{Particles}
We use $\sim$12 million macro-particles, each representing a swarm of real particles with a given size and mass. The initial spatial distribution of macro-particles is proportional to the local gas density, with particle sizes ranging from 10 $\mu$m to 1 cm, and a constant number of macro-particles per logarithmic size bin. Rather than making assumptions about the settling and actual size distribution, we instead analyse sub-populations of our initial distribution. Specifically, to measure accretion rate as a function of particle size, we tag and follow only the particles that initially reside within one Hill radius of the midplane. See PNRO18 for more details about the particle distribution, their motion, \rev{particle size distribution within macro-particles}, and selection.

\section{Results and Discussion}
\label{sec:results}

\subsection {Gas dynamics}
Figure \ref{fig:temp_rho_shell} shows the thermal and density structures of the envelopes for the five cases considered.
\begin{figure*}
  \centering
  \includegraphics[width=0.32\textwidth]{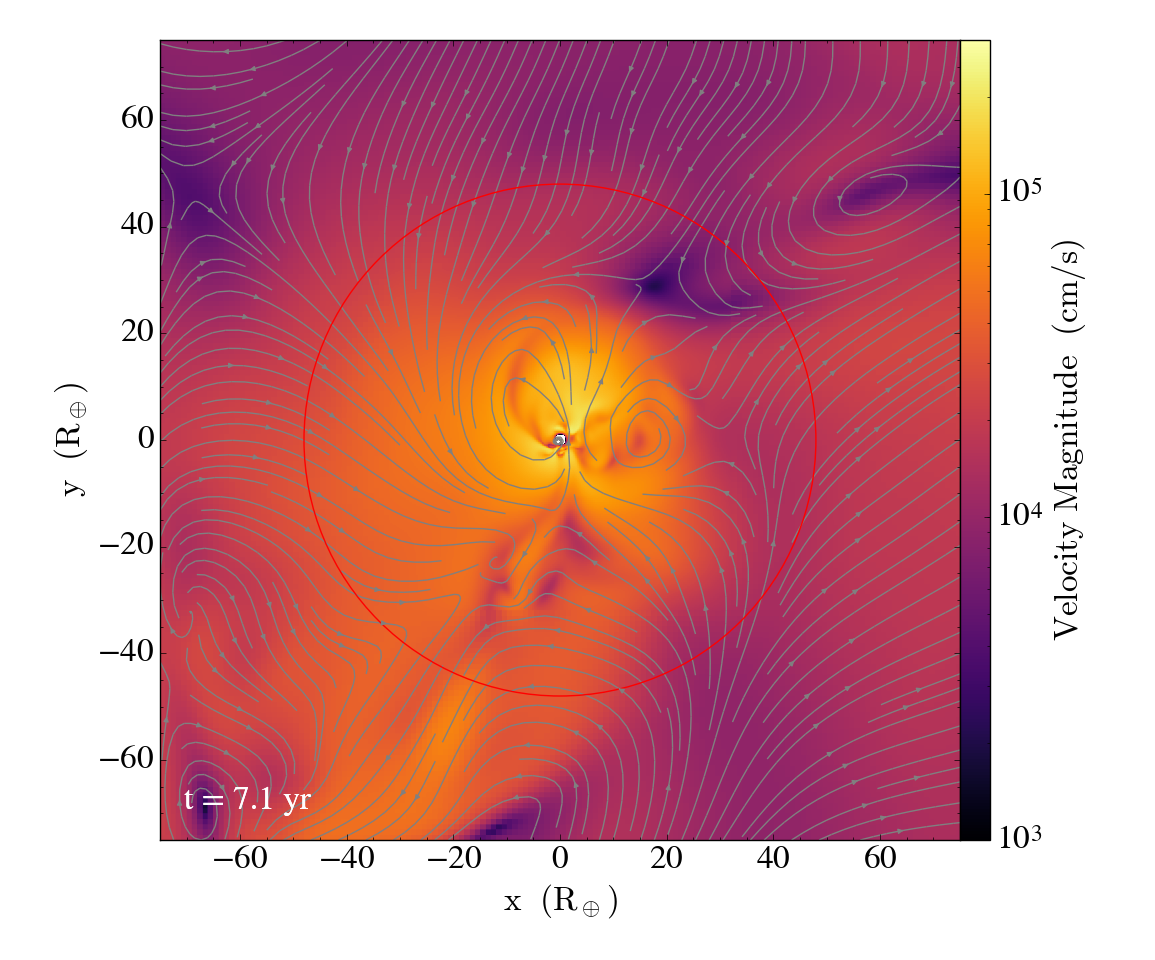}
  \includegraphics[width=0.32\textwidth]{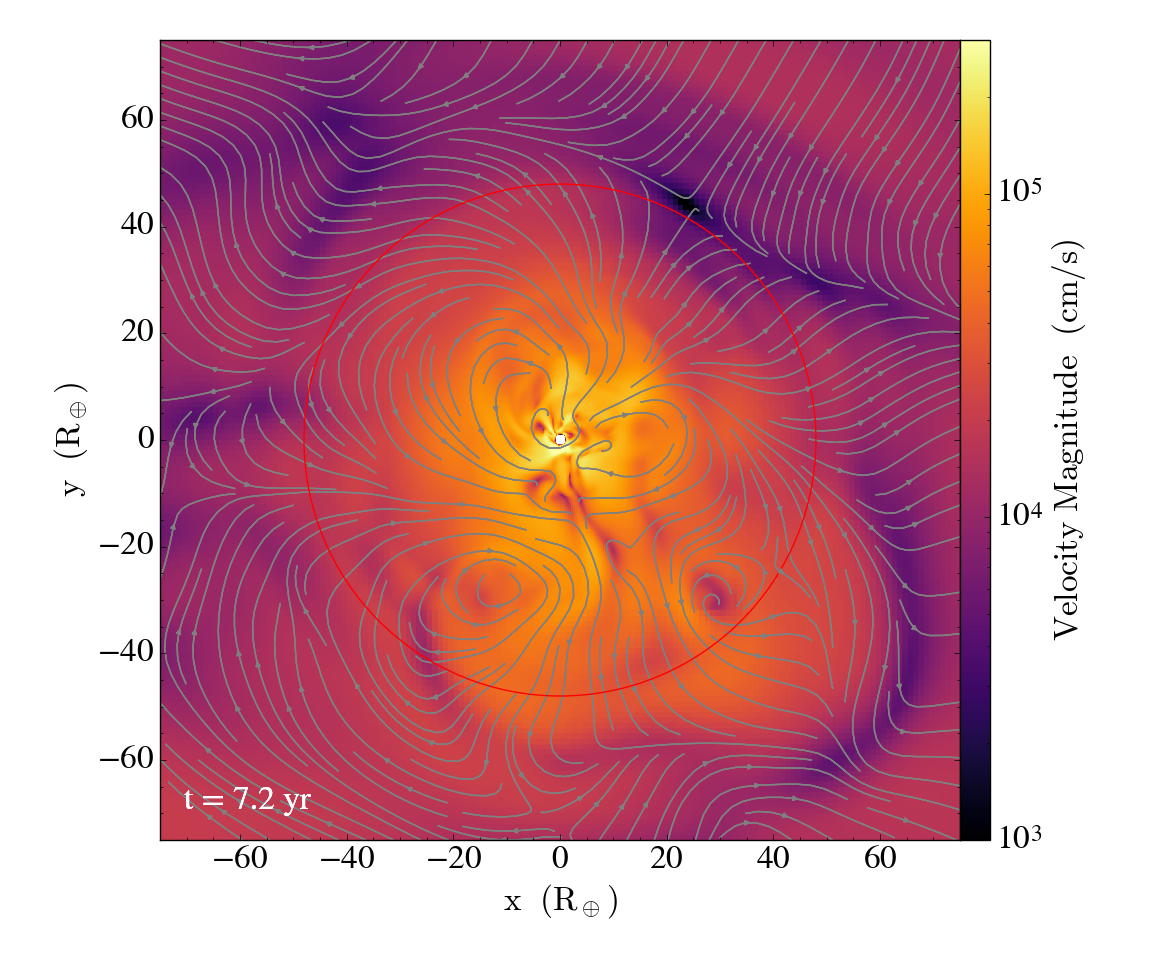}
  \includegraphics[width=0.32\textwidth]{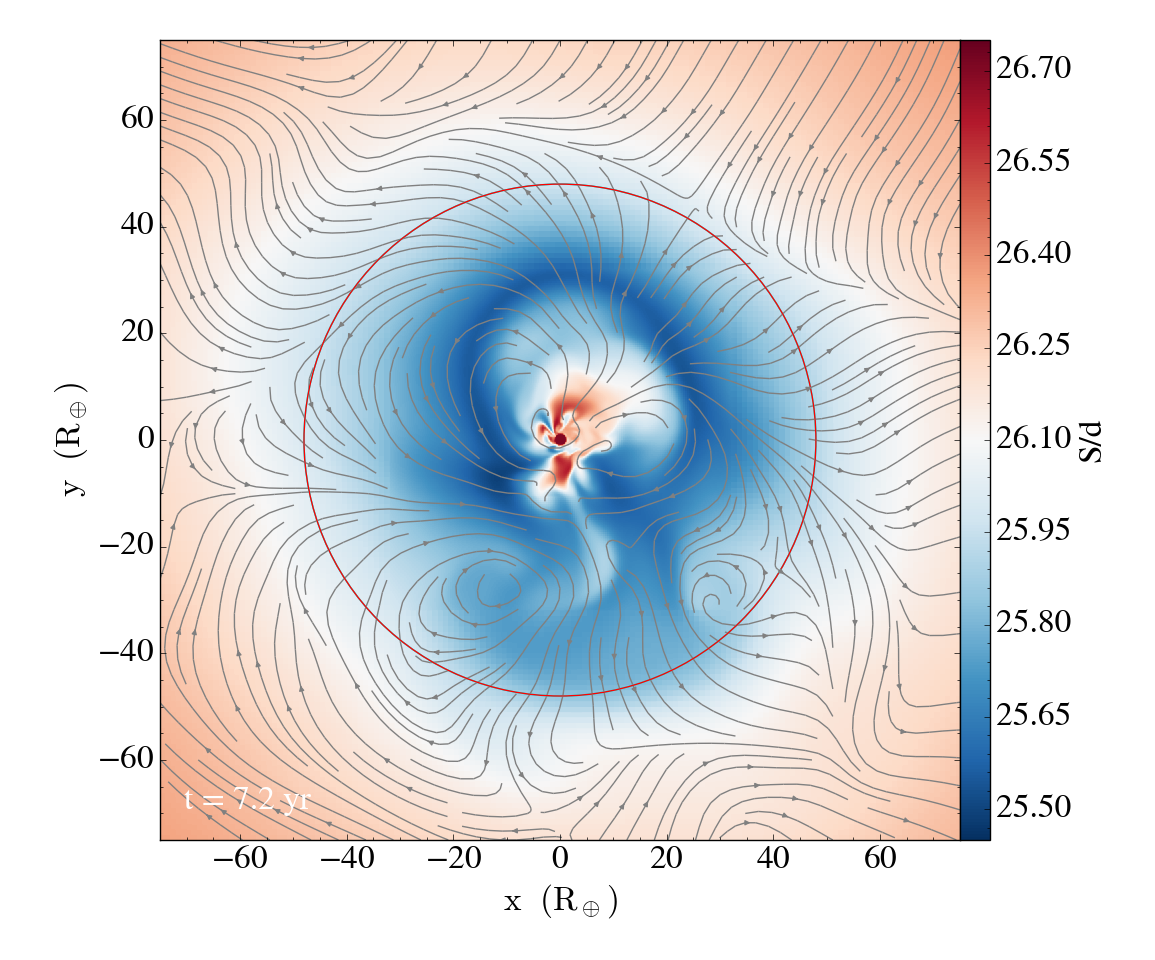}
  \caption{\textit{Left:} Velocity magnitude in the midplane in the \texttt{m095-conv-2e-6} simulation. \textit{Middle:} The same for the \texttt{m095-conv-2e-6-rt} simulation. \textit{Right:} Entropy per unit mass in the midplane in the \texttt{m095-conv-2e-6-rt} simulation. The red circles denote the canonical Bondi sphere. Gray streamlines indicate gas flow patterns (projected onto the midplane).}
  \label{fig:midplane_vel_sd_xy}
\end{figure*}
Convective motions effectively transport the excess heating from accretion of solids\rev{,} even when RT is not considered (or $\kappa$ is sufficiently high), and therefore the differences in thermal structure relative to the purely convective case are small. When radiative energy transport is \rev{used, radiative cooling effects become important in the outer parts of the atmosphere and the temperature decreases. However, cooling effects do not penetrate all the way to the embryo, and} the thermal structure near the embryo remains close to the adiabatic one. As the envelope adjusts its near-hydrostatic equilibrium to the lower temperature, it becomes on average denser.  Since this is an effect that accumulates \rev{with} depth---pressure scale height \rev{is} proportional to the local temperature---the effect on the density is larger than on the temperature. \rev{Indeed,} in the radiatively-cooled models, the total mass of the atmosphere is increased by about 4\% and 27\% for $\dot{M}$= 2\ 10$^{-5}$ and 2\ 10$^{-6}$ \rev{M$_\oplus$\! yr$^{-1}$}, respectively \rev{(with $\kappa$ = 0.1 cm$^{2}$\! g\rev{$^{-1}$})}, relative to adiabatic and purely convective models.

\begin{figure}
\centering
    \includegraphics[width=\columnwidth,clip,trim=0 10 0 0]{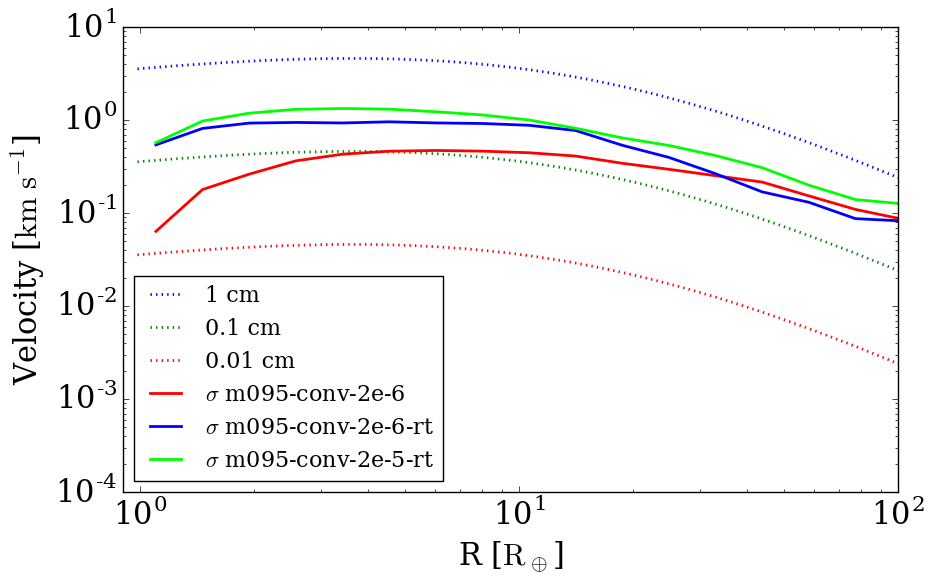}
    \caption{Time-averaged velocity dispersion \rev{($\sigma$)} profiles \rev{(solid lines)} for simulations \texttt{m095-conv-2e-6} (\rev{red}), \texttt{m095-conv-2e-6-rt} (\rev{blue}), and \texttt{m095-conv-2e-5-rt} (\rev{green}), shown against a background of free fall drag velocities \rev{(dotted)} of particles of size 1 cm (blue), 0.1 cm (green), and 0.01 cm (red). \rev{The velocity dispersions of \texttt{m095-conv-2e-6-rt-$\kappa$1} and \texttt{m095-conv-2e-6} simulations are very similar, and thus the former is omitted for clarity.}}
    \label{fig:vel_disp}
\end{figure}

Figure \ref{fig:midplane_vel_sd_xy} \rev{(middle)} shows an example of the convective velocity flow for a simulation where radiative energy transfer is included (\texttt{m095-conv-2e-6-rt}). The \rev{fluid motions} are similar \rev{to} the case with pure convection \rev{(left panel of Fig.\ \ref{fig:midplane_vel_sd_xy})}, but the velocity amplitudes are larger---especially at small radii---when radiative energy transfer is included. The reason for the increased velocity amplitudes is illustrated in \rev{the right panel of} Fig.\ \ref{fig:midplane_vel_sd_xy}. The radiative cooling tends to lower the entropy in the neighbourhood of the embryo \rev{relative to runs without RT}, while accretion heating \rev{simultaneously} keeps adding entropy, and the temperature of the embryo surface is held fixed. The net effect is a tendency to increase the temperature gradient, thus making it more super-adiabatic than in the case with only accretion heating, resulting in stronger driving of the convection. Near the embryo, the increase of the convective energy transport is largely able to compensate for the increased radiative cooling. Further out, where the convection motions are weaker and the \rev{vertical} optical depth \rev{is} smaller\rev{,} the effects of radiative cooling become more visible.

The increase of convective velocity amplitudes is illustrated in Fig.\ \ref{fig:vel_disp} against a background of particle drift speeds. The figure shows that velocity amplitudes near the embryo \rev{for simulations that include RT} are increased \rev{by} nearly an order of magnitude \rev{over the simulation that does not}, becoming comparable to the vertical drift speed of 2-3 mm particles. \rev{Conversely, however, increasing the accretion heating further does not significantly increase the dispersion.} Further out, where the horseshoe and shear flow patterns start to contribute to the velocity dispersion, the effects of radiative cooling on the velocity dispersion diminish.

\vspace*{-1cm}
\subsection {Solid accretion rates}

\begin{figure}
\centering
    \includegraphics[width=\columnwidth,clip,trim=0 13 0 10]{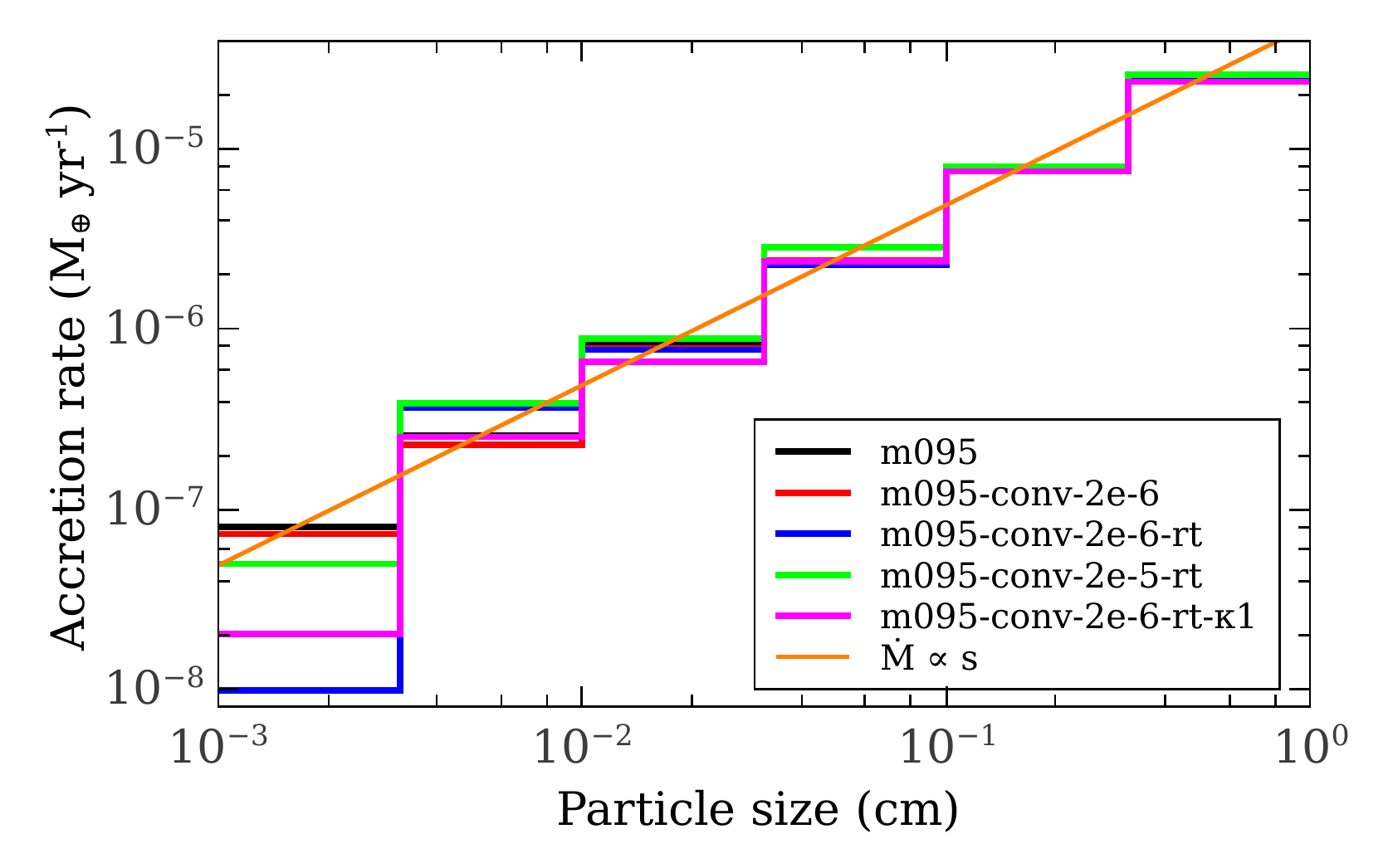}
    \caption{Pebble accretion rates for particle sizes within each size bin, drawn from a distribution of particles initially within $\pm R_\mathrm{H}$ of the midplane. The orange line shows a mass accretion rate proportional to particle size, $s$.}
    \label{fig:acc_rates}
\end{figure}

In \rev{our} previous paper (PNRO18), we determined the accretion rates of solids and showed that they scale linearly with the particle size. In Figure \ref{fig:acc_rates}, we affirm that the previously determined accretion rates are robust. Indeed, the accretion rates in simulation runs with convective motions and radiative cooling do not differ significantly from runs without; deviations are mainly due to increased noise as the convective motions of the gas stir the solids. \rev{Moreover, we find that stronger convective motions due to a larger accretion heating (e.g.\ that could result from changing the particle size distribution) do not significantly affect the accretion rates. Although the amplitude of the convective motions increase with accretion heating, they remain confined to a region $\lesssim 30 R_{\rm p}$ (Fig.\ \ref{fig:vel_disp}). Whether a particle will be accreted or not is, meanwhile, determined at larger radii (cf.\ Fig.\ 19 of PNRO18), and thus the strength of the convection does not significantly affect the accretion rate of particles.}

\section{Conclusions and Outlook}
\label{sec:conclusions}
In these first-of-a-kind, three-dimensional, radiative-convective models of hot and extended primordial atmospheres of Earth-mass embryos, with realistic---albeit schematic---scattering included, we find that radiative cooling \rev{leads to a} local increase of the shell averaged mass density on the order of a factor of two \rev{, but only a slight decrease in the temperature}. For the conditions modelled here, cooling effects do not penetrate to the bottom of the atmosphere where convection is the dominant mechanism of energy transport. In response to the tendency from cooling to increase the radial temperature gradient, however, the amplitude of convective motions increase by nearly an order of magnitude near the embryo surface relative to runs without RT. 

Even though radiative cooling \rev{modifies the atmosphere structure}, the accretion heating and resulting convective energy transport still dominate near the planetary embryo, where the atmosphere temperature remains ``anchored'' to the surface temperature of the embryo. This is encouraging, since it implies that planetary embryos embedded in protoplanetary disks can retain hot atmospheres \rev{by adjusting their nearly adiabatic stratifications only slightly,} throughout much of the evolution of the disk. For cooling to become important at earlier times\rev{,} the opacities would have to be much lower than assumed here (0.1 - 1 g\! cm$^{-2}$).

When the disk and primordial atmosphere finally become genuinely optically thin, the atmospheres that remain around low mass planetary embryos are therefore relatively light, with their future fate depending on the relative balance between the slow cooling of the still hot embryo, and the secular loss of the remaining atmosphere. As argued by \citet{Ginzburg2016}, it is the outcome of this balance that ultimately determines if a planet ends up as a \rev{gas-rich planet} with a significant H+He atmosphere, or becomes a rocky planet, with an atmosphere consisting of heavier gas molecules, possibly even dominated by out-gassing.

The current modelling should be seen as a pilot effort, exploring effects and probing what is currently possible. \rev{As briefly discussed in Appendix \ref{app:tech_con} (available online only), the current results were obtained with a moderate computational effort.} Obvious factors that could use improvement are the EOS, the constant and grey opacity, and the spatial resolution, which could stand to be increased in order to better resolve the convective motions. Improving all these factors is certainly possible\rev{, and will be the subject of future work}.
\vspace*{-0.5cm}
\section*{Acknowledgements}
We  are grateful to the anonymous reviewer for a constructive report that helped to significantly improve the quality of this manuscript. The work of AP and \AA N was supported by grant 1323-00199B from the Danish Council for Independent Research (DFF). The Centre for Star and Planet Formation is funded by the Danish National Research Foundation (DNRF97). Storage and computing resources at the University of Copenhagen HPC centre, funded in part by the Villum Foundation (VKR023406), were used to carry out the simulations.


\vspace*{-0.5cm}
\bibliographystyle{mnras}
\bibliography{references} 



\appendix
\section{The ray-based radiation solver}
\label{app:ray_tracing}

Being a 7-dimensional problem (three space + time + wavelength + unit direction vector (polar and azimuthal angles), RT can be a daunting physical process to solve. A number of approaches have been developed over the last several decades in an attempt to tackle the problem. One common approach is to reduce the dimensionality of the problem. These approximations are known as moment methods, since they take the moments of radiative transfer equation (i.e.\ the angular dependence is averaged out). It can be a good approximation for smooth, diffuse radiation fields in optically thick media when the radiation is tightly coupled to the gas. Common approximations of various complexity include flux limited diffusion (FLD) (e.g. \citealt{levermorepomraning1981_fld,gonzalez2015}), the M1 method (e.g. \citealt{Rosdahl2015}), or variable tensor Eddington factors \citep{dullemondturolla2000_vtef}. Although protoplanetary disks and the deep interiors of the planetary atmospheres are generally optically thick, there are always radiative zones where moment methods are not reliable.

An alternative to moment methods are ray-tracing techniques. In this case, the accuracy of the general solution depends on the sampling of ray directions and the density of sampling points. Naturally, more directions and greater sampling density means the RT solver becomes more accurate but also more computationally expensive. In a short-characteristics ray-tracing scheme (e.g.\ \citealt{stonetal1992}), only the neighbouring grid cells are used to interpolate the intensities. It is faster than other ray-based approaches, but is also more diffusive. In a long-characteristics scheme (e.g.\ \citealt{nordlund82,Heinemann2006}), rays are traced cell-by-cell across long distances. This scheme provides a maximum possible accuracy at the cost of a large number of redundant calculations. Hybrid-characteristics ray-tracing schemes \citep{rijhorstetal2006_hybrid} are a combination of long and short characteristics methods. In addition to moment methods and ray-tracing, it is of course also possible to solve radiative transfer using Fourier transforms (e.g.\ \citealt{cen2002}) or Monte Carlo techniques (e.g.\ \citealt{robitaille2011}).

Although briefly described in \citep{Nordlund2018}, here we present additional details about the hybrid-characteristics ray-tracing RT scheme that we employ in the simulations presented here. In DISPATCH, the RT scheme uses long rays inside patches and short rays in-between. The RT solver consists of a ray geometry component, an initialization component, and runtime schemes. In addition, since the RT depends on the values of, for example, density and temperature, it relies on the co-existence of instances of (magneto-)hydrodynamical patches that can provide these quantities.
\subsection{The equation of radiative transfer}
\label{subsec:transfer_eq}
The radiative transfer equation describes the change of the specific intensity during its propagation through a medium and is determined by emission and absorption processes. Solving the equation of radiative energy transport (e.g. \citealt{Mihalas1984}) gives the net heating or cooling,
\begin{equation} 
Q = \oint_{4\pi} d\Omega_\angle \int_0^\infty d\nu \rho \kappa_\nu \left(I_{\Omega_\angle,\nu}-S_{\nu}\right),
\label{eq:transfer_eq}
\end{equation}
which can then be coupled to the fluid dynamics. In Eq.\ \ref{eq:transfer_eq}, $\Omega_\angle$ is a solid angle, $\nu$ is frequency, $\rho$ is the mass density, $\kappa_\nu$ is the absorption coefficient per unit frequency, $I_{\Omega_\angle,\nu}$ and $S_{\nu}$ are the specific intensity and the source function per unit area, frequency, time, and solid angle. It is advantageous to write the radiative transfer equation directly in the net difference $I-S$ rather than in terms of $I$, as is typical, to circumvent numerical round-off problems where the optical depth is very high and $I$ approaches closely to $S$ \citep{nordlund82}. At large optical depth, the difference $I-S$ goes as the 2$^{\rm{nd}}$ order derivative of the source function, and therefore formulations that assume constant or linear $S$ within an interval are not accurate enough. In this work, we therefore use an integral formulation which represents $S$ to 2$^{\rm{nd}}$ order. A modified Feautrier solver following \cite{nordlund82} is, however, also available in the code.

\subsubsection{Integral formulation}
Given an optical depth scale $\tau_\nu = \int \rho \kappa_\nu ds$, where $s$ is the optical distance along the propagation direction of the ray, Eq.\ \ref{eq:transfer_eq} can be rewritten as
\begin{equation}
\frac{dQ^+}{d\tau} = Q^+ - \frac{dS}{d\tau}
\label{eq:int_fwd}
\end{equation}
in the direction of increasing $\tau$ and
\begin{equation}
\frac{dQ^-}{d\tau} =-Q^- + \frac{dS}{d\tau}
\label{eq:int_rev}
\end{equation}
in the direction of decreasing $\tau$. For simplicity, the explicit reference to the frequency $\nu$ has been dropped here. Equations \ref{eq:int_fwd} and \ref{eq:int_rev} have the same form as Eq.\ \ref{eq:transfer_eq}, but with a term $\frac{dS}{d\tau}$ rather than $S$. For an assumed quadratic source function $S$, the source function derivative is a linear function, and it is therefore easy to establish an analytical solution from one point to the next:
\begin{align}
Q^+(\tau_i) & = Q^+(\tau_{i-1})a_i - S'(\tau_{i})b_i + S''(\tau_{i})c_i;\\
Q^-(\tau_{i}) & = Q^-(\tau_{i+1})a_{i+1} + S'(\tau_{i})b_{i+1} + S''(\tau_{i})c_{i+1},
\end{align}
where the coefficients $a_i$, $b_i$ and $c_i$ are:
\begin{equation}
\begin{aligned}
a_i &= e^{-\Delta \tau} \\
b_i &= 1-a_i \\
c_i &= b_i-\Delta\tau a_i,
\end{aligned}
\end{equation}
and $\Delta \tau = \tau_i -\tau_{i-1}$.

\subsection{Ray geometry}
\label{subsec:ray_geometry}
When the RT module is initialized, it first creates a \textit{ray geometry} (RG) for existing patches. The only information the RG component needs is the simulation geometry type (i.e.\ Cartesian/cylindrical/spherical), dimensions of the patch (number of cells per direction), the desired number of ray-directions, and their angular separation.
\subsubsection{Ray tracing through a patch}
\begin{figure}
  \includegraphics[width=0.9\columnwidth]{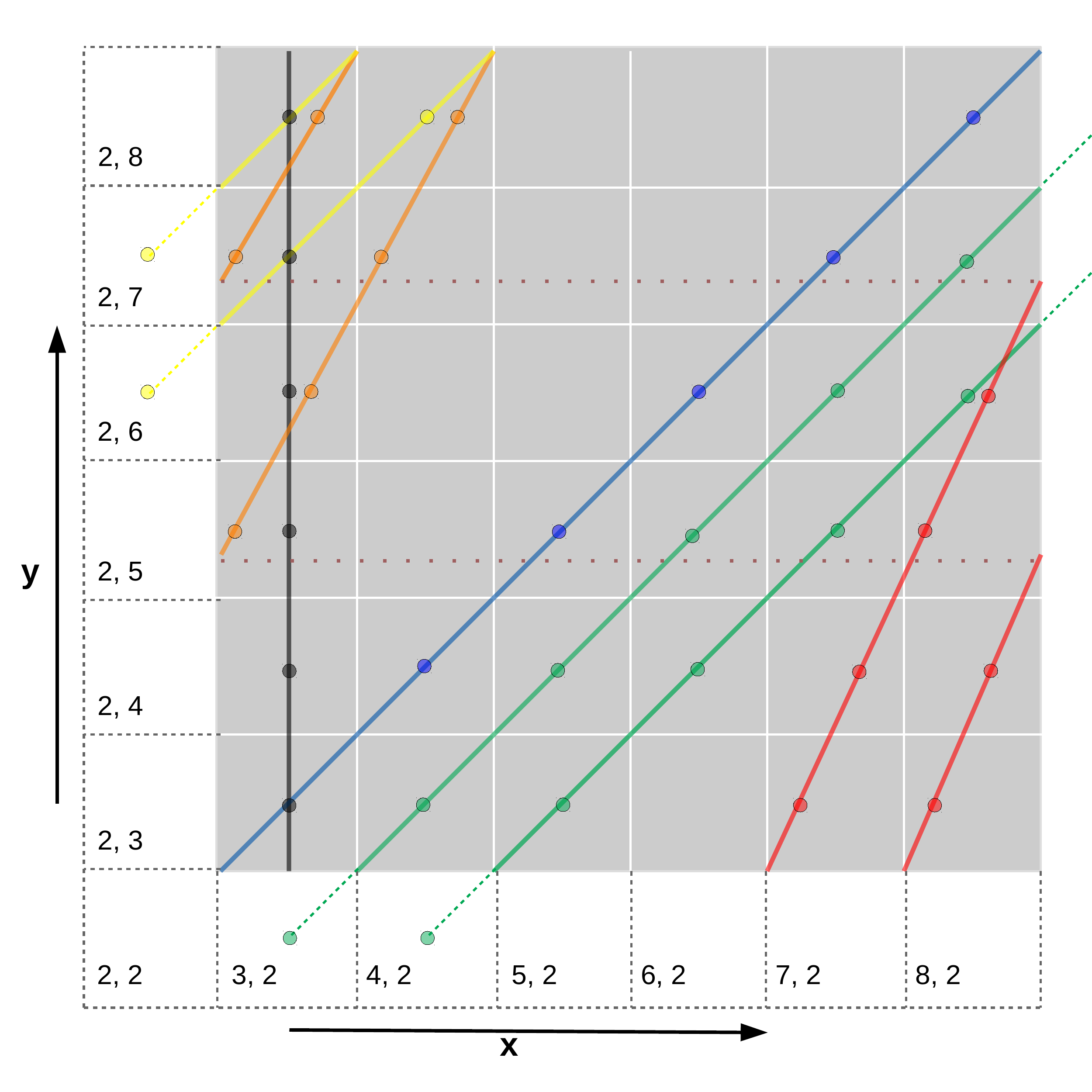}
  \caption{ Ray geometry on a 2D patch. Cells are gray with white borders, guard cells are white with dashed black borders. Different ray colours represent different directions, and circles represent ray points where the hydrodynamic quantities are interpolated. Typically, we employ two guard cells, but only the innermost guard cell is shown here. Therefore, the numbering starts at two. See the text for additional details.}
  \label{fig:rt_grid}
\end{figure}

The ray casting starts by selecting the \textit{main} direction; this is the global coordinate that forms the smallest angle with the ray. If the main direction is along, e.g., the $z$-axis, the rays are cast from the lower $xy$-plane with a ray spacing in the $x$- or $y$-direction equal to the patch cell size in that direction. The 45$^\circ$ angles are the exception -- if they are used, the selection of the \textit{main} direction is hard-coded to $x$- and $y$-directions, thus avoiding repetition. Ray points, where the required hydrodynamic quantities are interpolated to the ray, are cell-centred along the direction of the ray. With such positioning, one needs to use 2D interpolations for a 3D space. Ideally, the ray points are coincident with patch cell centres. In this case, the required hydrodynamic quantities (e.g.\ density and temperature) do not need to be interpolated to the RG, and subsequently the resulting heating rates neither need to be interpolated back to the mesh, saving a significant amount of computational time. Figure \ref{fig:rt_grid} presents a 2D representation of the RG. Rays with three different inclinations are shown; circles denote where the hydrodynamic data is interpolated from.

To trace a ray through a patch we use a scheme that is similar to the ``fast voxel traversal algorithm'' \citep{Amanatides87}, but is slightly simpler, and therefore a bit faster:
\begin{enumerate}
\item Consider a ray with a \textit{main} direction along the $z$ axis. First, calculate the distance $r$ the ray travels to cross one cell along the \textit{main} axis:
\begin{equation}
r = \frac{ds_z}{\cos\theta},
\end{equation}
where $\theta$ is the polar angle in the zenith direction and $ds_z$ is the cell size in the same direction.
\item Next, calculate a step size in the $x$- and $y$-directions that the ray must take in order to travel the distance $r$:
\begin{equation}
dp_x = r \sin\theta \cos\phi
\end{equation}
and
\begin{equation}
dp_y = r \sin\theta \sin\phi.
\end{equation}
The step size in the $z$-direction is simply $ds_z$. 
\item Step back to the nearest guard cell. Rays need to know exactly where in an upstream patch or boundary the incoming heating rates should be taken from. For example, in Figure \ref{fig:rt_grid}, if we consider the green ray, which starts in cell $[4,3]$, the upstream point is the guard cell $[4,2]$, as indicated by the dashed green line. This is stored as the upstream point for this ray in the forward direction.
\item Starting from the first point (e.g.\ $[4,3]$ in Figure \ref{fig:rt_grid}), we enter a loop over the position $p$ and simply add the step size:
\begin{equation}
p_{x,y,z} = p_{x,y,z} + dp_{x,y,z}.
\end{equation}
\item After each step, check if the ray is still inside the patch (i.e.\ it has not crossed any of the patch boundaries). If not, continue the loop. 
\item However, if the ray has left the patch after the last step, store the last position as the downstream point for this ray in the forward direction (or as an upstream point in the reverse direction). Depending on which wall was crossed, we either:
\begin{itemize}
\item stop the ray if the $z$-boundary was crossed, or
\item restart the ray. This is done by resetting the index, which is currently outside the patch, to the lower inner value (e.g.\ in Figure \ref{fig:rt_grid}, the yellow rays are restarted from the green rays while the orange rays are restarted from the red ones) and going back to step (iii).
\end{itemize}
\end{enumerate}
The calculations in steps (i) and (ii) are done once per direction. This procedure requires only three floating point additions and three floating point comparisons per iteration. 

\subsubsection{Hierarchy}
\label{subsubsec:hierarchy}
Once the casting of all the rays inside a patch is finished, we have a large stack of rays with different lengths, that have crossed different walls, and are in different directions. The results are now sorted following a particular hierarchy:
\begin{enumerate}
\item \textit{Ray directions} -- in principle, as long as scattering is not considered, rays in one direction do not need to know about rays in another direction, and they can be updated as independent tasks by different threads. To solve RT inside a patch, boundary conditions from the patch wall that a ray originates from are required (i.e.\ the upstream/downstream points that were stored during the ray tracing procedure). Note that not all slanted rays with one direction originate and terminate at the same wall (see Figure \ref{fig:rt_grid}).
\item The complete set of rays is further sub-divided into \textit{ray bundles}; these are defined as sets of rays that originate and terminate on the same pair of patch walls. This helps avoid waiting time originating from sets of ray directions that end on more than one patch wall. Ray bundles also hold meta-data about the rays it contains: their step sizes $dp_{xyz}$, their spatial orientation (the main direction and inclination angles), and how the rays are sorted within the bundle. Ray-based RT is, in general, a repetitive task, particularly when a large number of rays is considered. It is thus highly advantageous to use schemes that maximize vectorization and therefore computational efficiency. As such, all rays should, preferably, be the same length. This is a natural feature of rays that are parallel to a patch coordinate axis.
\item Slanted rays, meanwhile, are further rearranged into \textit{ray packets}; these are sets of rays in a ray bundle that have the same length. By organising the ray packet as a single array, we promote faster data lookup (e.g.\ interpolating ray coordinates to patch coordinates, origin/termination points, etc.) and vectorization.
\end{enumerate}

In this work, we use 26 ray directions with 45$^{\circ}$ angular separation between rays. These rays have the advantage that they will generally terminate at locations coincident with the origins of rays in a `downstream' (with respect to RT) patch. This feature is demonstrated in Figure \ref{fig:rt_grid}, where red rays terminate at the $+x$ wall and then restart at $-x$ wall (yellow rays), with exactly the same $y$ position.

Finally, we note that ``bundle chains'' are not used in the RT in this work (in contrast to \citealt{Nordlund2018}). We, instead, run the RT solver in so-called ``no-chains'' mode. 

\subsubsection{Initialisation and execution}
After the ray geometry has been generated, the RT initialization component iterates through all of the hydrodynamical patches in the task list and creates a corresponding and overlapping \textit{RT patch}. A single RT patch keeps track of all the ray-directions crossing the patch. For each ray bundle, the RT patch maintains a corresponding array (with size equal to the total number of rays in this bundle) which points to the appropriate locations in the guard cells where rays must pick up the necessary upstream/downstream values. 

Now that all the RT patches and rays are set up, the initialization procedure generates neighbour relations between the RT patches similarly to how it is done for hydrodynamical patches \citep[see][]{Nordlund2018}.

Finally, RT is ready for execution. Courtesy of the connection to an underling hydrodynamical patch, the RT patch is updated immediately after its ``parent'' hydrodynamical patch. After the mean intensities everywhere within the patch are updated, the heating rates (Eq.\ \ref{eq:rt_qheating}) are calculated and stored for use by the underlying hydrodynamical patch.

\subsubsection{Applying heating/cooling rates to hydrodynamics}
Once the RT solver returns the frequency-dependent heating rates, they are stored, via reverse volume mapping, on a mesh similar to the one belonging to the parent hydrodynamical patch. $Q$ is then integrated over frequency bins and the different directions. At this point, the heating rate $Q$ is ready to be applied to the hydrodynamics. Whenever the hydrodynamical patch needs the heating rates, it picks them up directly without need for further interpolation.


\section{Performance considerations}
\label{app:tech_con}

The current results in this work were obtained with only moderate computational effort, with each model requiring of the order of a few thousand core hours. The DISPATCH code framework \citep{Nordlund2018} can handle arbitrary equations of state and opacities using an optimized table lookup without significantly increasing the computational cost. The RT solver is also ready for multi-frequency experiments, using either representative frequencies or opacity distribution functions. Increasing the spatial resolution results, as always, in a cost increase scaling with the inverse fourth order power of the smallest resolution element. Although this can be partly mitigated by the use of local time steps (\rev{e.g.\ by splitting a spatial region into a number of patches each with their own time step}), the scaling of the computing \rev{cost} will remain essentially $N_{\rm bin} N_{\rm cell}^{4/3}$\rev{, where $N_{\rm bin}$ is the number of opacity bins}. Taking advantage of the linear weak scaling of DISPATCH \citep[cf.][Fig.\ 5]{Nordlund2018}, it is thus possible to perform simulations with a tabular EOS and opacities, several opacity bins, and resolution increased by a factor of 8, at costs per model of the order 10-20 million core hours. Alternatively one could run, using similar amounts of computing resources, a dozen or more models with a factor of four better resolution than the current \rev{endeavour}.

\bsp	
\label{lastpage}
\end{document}